\documentclass[aps,twocolumn,prl,tightenlines,floatfix]{revtex4-1}

\usepackage[dvips]{graphicx}
\usepackage[english]{babel}
\usepackage{amsmath}
\usepackage{amssymb}
\usepackage{times}

\begin{document}

\title{Enhancement effect of mass imbalance on 
  Fulde-Ferrell-Larkin-Ovchinnikov type of pairing in Fermi-Fermi
  mixtures of ultracold quantum gases}

\author{Jibiao Wang, Yanming Che, Leifeng Zhang }
\affiliation{Department of Physics and Zhejiang Institute of Modern
  Physics, Zhejiang University, Hangzhou, Zhejiang 310027, China and Synergetic Innovation Center of Quantum Information and
  Quantum Physics, Hefei, Anhui 230026, China} 
\author{Qijin Chen}
\email[Corresponding author: ]{qchen@zju.edu.cn}
\affiliation{Department of Physics and Zhejiang Institute of Modern
  Physics, Zhejiang University, Hangzhou, Zhejiang 310027, China and Synergetic Innovation Center of Quantum Information and
  Quantum Physics, Hefei, Anhui 230026, China}

\date{\today}

\begin{abstract}
  Ultracold two-component Fermi gases with a tunable population
  imbalance have provided an excellent opportunity for studying the
  exotic Fulde-Ferrell-Larkin-Ovchinnikov (FFLO) states, which have
  been of great interest in condensed matter physics. However, the
  FFLO states have not been observed experimentally in Fermi gases in
  three dimensions (3D), possibly due to their small phase space
  volume and extremely low temperature required for an equal-mass
  Fermi gas.  Here we explore possible effects of mass imbalance,
  mainly in a $^6$Li--$^{40}$K mixture, on the one-plane-wave FFLO
  phases for a 3D homogeneous case at the mean-field level.  We
  present various phase diagrams related to the FFLO states at both
  zero and finite temperatures, throughout the BCS-BEC crossover, and
  show that a large mass ratio may enhance substantially FFLO type of
  pairing.
\end{abstract}

\pacs{03.75.Ss,03.75.Hh,67.85.-d,74.25.Dw \hfill \textsf{arXiv:1404.5696}}

\maketitle

The past decade has seen great progress in ultracold atomic Fermi gas
studies \cite{Review,Bloch_RMP}. With the easy tunability in terms of
interaction, dimensionality, population imbalance as well as mass
imbalance \cite{Review,Bloch_RMP}, ultracold Fermi gases have provided
a good opportunity to study many exotic quantum phenomena. In
particular, the Fulde-Ferrell-Larkin-Ovchinnikov (FFLO) states, which
were first predicted by Fulde and Ferrell \cite{FF} (FF) and Larkin
and Ovchinnikov \cite{LO} (LO) in an $s$-wave superconductor in the
presence of a Zeeman field about fifty years ago, have attracted
enormous attention in condensed matter physics \cite{LOFF_Review},
including heavy-fermion \cite{BianchiPRL89},
%,Radovan,Kenzelmann},
organic \cite{LebedPRB82} %ShimaharaJPSJ71,
 and high $T_c$
superconductors
\cite{GrafPRB72,*SimonsPRL102,*ChoPRA83},
%,TingPRL96,Ptok2013
nuclear matter \cite{SedrakianPRC67} and color superconductivity
\cite{Alford}, %,Bowers
and ultracold Fermi gases \cite{SedrakianPRA72,*Liao,*Sheehy_RPP}.  In
these exotic states, Cooper pairs condense at a finite momentum
$\textbf{q}$, with an order parameter of the form of either a
plane-wave
$\Delta(\textbf{r})=\Delta_{0}e^{i\textbf{q}\cdot\textbf{r}}$ or a
standing wave
$\Delta(\textbf{r})=\Delta_{0}\cos(\textbf{q}\centerdot\textbf{r})$
for the FF and LO states, respectively.  Despite many theoretical
studies on the FFLO states in equal-mass Fermi gases, both in a 3D
homogeneous case
\cite{LOFF1,*SR06,*huhui06pra,helianyi06prb,*CombescotPRB7114,*Yip07pra}
and in a trap \cite{Machida2,zw07pra,Kinnunen}, the experimental
search for these exotic states in atomic Fermi gases still has not
been successful \cite{ZSSK06,*Rice1,Shin08}, largely because they
exist only in a \emph{small} region at \emph{very low} temperature in
the phase space \cite{LOFF1,*SR06,*huhui06pra,zw07pra}. To find these
elusive states, attention has been paid to more complex systems. There
have been theoretical investigations in either Fermi-Fermi mixtures
\cite{Stoof09prl,*Stoof10pra,Stoof12pra} or equal-mass Fermi gases
with spin-orbit coupling
\cite{chuanwei12,*xiaji13,Yi2013PRA87,*Yi2013PRL,*Pu2013,*HuLiu2013NJP,*Iskin2013}
or in an optical lattice
\cite{WuCJPRA83,*Buchleitner2012,*Mendoza2013,*Koga2012,*Torma2012,*ChenAHai2012}.
Recently, Stoof and coworkers \cite{Stoof09prl,*Stoof10pra} found an
instability toward a supersolid state \cite{footnote1} in a
homogeneous $^6$Li--$^{40}$K mixture in the unitarity and BCS
regimes. Using a mean-field theory and the Bogoliubov--de Gennes (BdG)
formalism, they have also studied the LO states for the unitary case
\cite{Stoof12pra}. However, it is hard to perform stability analysis
for various phases in the BdG formalism. Other types of
mass-imbalanced systems such as the $^6$Li-$^{173}$Yb mixture were not
considered in Ref.~\cite{Stoof12pra}.

In this paper, we will investigate the \emph{one-plane-wave} FFLO
states (i.e., the FF states) in a homogeneous $^6$Li--$^{40}$K
mixture, as well as for other mass ratios, as they undergo BCS--BEC
crossover, using a mean-field theory. We will present several $T$--$p$
(where $p$ is the population imbalance) phase diagrams to show the
FFLO regions under typical interaction strengths ($1/k_{F}a$) at
finite temperature, as well as $p$--$1/k_{F}a$ phase diagrams at zero
temperature. We find that when the heavy species, $^{40}$K, is the
majority, an $s$-wave FFLO phase, which is stable against phase
separation, persists throughout the BCS through BEC regimes, with the
population imbalance evolving from small to large. In contrast, when
the light species, $^6$Li, is the majority, such an FFLO phase exists
only in the BCS regime.  At unitarity, the phase space of FFLO states
becomes substantially enlarged as the mass ratio increases.  The
superfluid transition temperature $T_c$ of the FFLO states may be
enhanced by a factor of about 3 and 7 for a large mass ratio as in
$^6$Li--$^{40}$K \cite{Grimm2008PRL,*Naik,*Dieckmann2008PRL} and
$^6$Li--$^{173}$Yb \cite{Fukuhara,Gupta2013PRA}, respectively, in
comparison with the equal-mass case \cite{LOFF1}, which is hardly
accessible experimentally \cite{Rice1}.  Therefore, one may find it
realistic to experimentally observe the exotic FFLO states in unitary
ultracold Fermi-Fermi mixtures with a large mass ratio.

We also find that at zero temperature the phase space 
of stable FFLO states becomes larger as the mass ratio increases.
This has never been reported by other previous works.

We consider a three-dimensional (3D) Fermi-Fermi mixture with a
short-range contact potential of strength $U<0$, where momentum
$\textbf{k}$ pairs with $\textbf{q}-\textbf{k}$ and thus Cooper pairs
have a nonzero center-of-mass momentum $\textbf{q}$.  The dispersion
of free atoms is given by $\xi_{\textbf{k},\sigma}=
\mathbf{k}^{2}/2m_{\sigma}-\mu_{\sigma}$, where $m_{\sigma}$ and
$\mu_{\sigma}$ are the mass and chemical potential for (pseudo)spin
$\sigma=\uparrow,\downarrow$, respectively. We set the volume $V=1$,
and $\hbar=k_{B}=1$. At the mean-field level, the system with a
one-plane wave LOFF solution can be described by the following
Hamiltonian
%\begin{equation}
%H=\sum_{\textbf{k},\sigma}\xi_{\textbf{k},\sigma}c^{\dagger}_{\textbf{k},\sigma}
%c_{\textbf{k},\sigma}+U\sum_{\textbf{k},\textbf{k}'}c^{\dagger}_{\textbf{k},\uparrow}
%c^{\dagger}_{\textbf{q}-\textbf{k},\downarrow}c_{\textbf{q}-\textbf{k}',\downarrow}
%c_{\textbf{k}',\uparrow}\,.
%\end{equation}
%After a mean-field treatment, the Hamiltonian can be rewritten as
%
\begin{eqnarray}
  H^{MF}&=&\sum_{\textbf{k},\sigma}\xi_{\textbf{k},\sigma}c^{\dagger}_{\textbf{k},\sigma}
  c_{\textbf{k},\sigma}\nonumber\\
  &&{}-\sum_{\textbf{k}}(\Delta c^{\dagger}_{\textbf{k},\uparrow}
  c^{\dagger}_{\textbf{q}-\textbf{k},\downarrow}+\Delta^{*}c_{\textbf{q}-\textbf{k},\downarrow}
  c_{\textbf{k},\uparrow})-\frac{\,\Delta^{2}}{U\,}\,,
\end{eqnarray}
where the order parameter $\Delta$ carries momentum $\mathbf{q}$.
Using Bogoliubov transformation, it is easy to deduce the gap equation
via the self-consistency condition
\begin{equation}
 \Delta\equiv-U\sum_{\textbf{k}}\langle c_{\textbf{k},\uparrow}
 c_{\textbf{q}-\textbf{k},\downarrow}\rangle\,,
\end{equation}
which can be written as
\begin{eqnarray}
0&=&\frac{1}{U}+\sum_{\textbf{k}}\left[\frac{1-f(E_{\textbf{kq},\uparrow})
 -f(E_{\textbf{kq},\downarrow})}{2E_{\textbf{kq}}}\right]\nonumber\\
 &=&\frac{1}{U}+\sum_{\textbf{k}}\frac{1-2\bar{f}(E_{\textbf{kq}})}{2E_{\textbf{kq}}}\,,
\end{eqnarray}
where $E_{\textbf{kq}}=\sqrt{\xi_{\textbf{kq}}^{2}+\Delta^{2}}$,
$E_{\textbf{kq},\uparrow}=E_{\textbf{kq}}+\zeta_{\textbf{kq}}$,
$E_{\textbf{kq},\downarrow}=E_{\textbf{kq}}-\zeta_{\textbf{kq}}$,
$\xi_{\textbf{kq}}=(\xi_{\textbf{k},\uparrow}
+\xi_{\textbf{q}-\textbf{k},\downarrow})/2$,
$\zeta_{\textbf{kq}}=(\xi_{\textbf{k},\uparrow}
-\xi_{\textbf{q}-\textbf{k},\downarrow})/2$.  We have defined the average
\begin{equation}
\bar{f}(x)\equiv[f(x+\zeta_{\textbf{kq}})+f(x-\zeta_{\textbf{kq}})]/2\,,
\end{equation}
where $f(x)$ is the Fermi distribution function. The coupling constant $U$
can be replaced by the dimensionless parameter, $1/k_{F}a$, via the 
Lippmann-Schwinger equation
\begin{equation}
 \frac{1}{U}=\frac{m_r}{4\pi a}-\sum_{\textbf{k}}\frac{1}{2\epsilon_{\textbf{k}}}\,,
\end{equation}
where $a$ is the $s$-wave scattering length,
$m_r=2m_{\uparrow}m_{\downarrow}/(m_{\uparrow}+m_{\downarrow})$ 
is twice the reduced mass, and $\epsilon_{\textbf{k}}=k^{2}/2m_r$.
Therefore the gap equation becomes
\begin{equation}
 \frac{m_r}{4\pi a}=\sum_{\textbf{k}}\Big[\frac{1}{2\epsilon_{\textbf{k}}}-
 \frac{1-2\bar{f}(E_{\textbf{kq}})}{2E_{\textbf{kq}}}\Big]\,.
 \label{eq:gap}
\end{equation}
The number density of each species is given by
\begin{equation}
 n_{\sigma}=\sum_{\textbf{k}}[f(E_{\textbf{kq},\sigma})u_{\textbf{kq}}^{2}
 +f(-E_{\textbf{kq},\bar{\sigma}})v_{\textbf{kq}}^{2}]\,,
\end{equation}
where $\bar{\sigma}=-\sigma$, and the coherence factors
$u_{\textbf{kq}}^{2}=(1+\xi_{\textbf{kq}}/E_{\textbf{kq}})/2$,
$v_{\textbf{kq}}^{2}=(1-\xi_{\textbf{kq}}/E_{\textbf{kq}})/2$.  So the
total number density $n=n_{\uparrow}+n_{\downarrow}$ and the density
difference $\delta n\equiv n_{\uparrow}-n_{\downarrow}$ are given by
\begin{eqnarray}
  n&=&\sum_{\textbf{k}}\Big[\Big(1-\frac{\xi_{\textbf{kq}}}{E_{\textbf{kq}}}\Big)+
  2\bar{f}(E_{\textbf{kq}})\frac{\xi_{\textbf{kq}}}{E_{\textbf{kq}}}\Big]\,,
  \label{eq:LOFF_neqa}\\
  \delta n&=&\sum_{\textbf{k}}
  \Big[f(E_{\textbf{kq},\uparrow})-f(E_{\textbf{kq},\downarrow})\Big]\,.
  \label{eq:LOFF_neqb}
\end{eqnarray}
The thermodynamic potential $\Omega_{S}$ is given by
\begin{equation}
 \Omega_{S}=-\frac{\,\Delta^{2}}{U\,} +
 \sum_{\textbf{k}}(\xi_{\textbf{kq}} -E_{\textbf{kq}})
 -T\sum_{\textbf{k},\sigma}\ln(1 + e^{-E_{\textbf{kq},\sigma}/T})\,.
\end{equation}
Momentum $\textbf{q}$ is determined by minimizing $\Omega_{S}$ at
$\textbf{q}$, i.e., $\frac{\partial\Omega_{S}}{\partial\textbf{q}}=0$,
which leads to
\begin{eqnarray}
\sum_{\mathbf{k}}\left[\frac{\mathbf{k}}
  {m_{\uparrow}}(n_{\textbf{kq}}+\delta n_{\textbf{kq}})+
  \frac{\mathbf{q-k}}{m_{\downarrow}}
  ( n_{\textbf{kq}}-\delta n_{\textbf{kq}})\right]=0\,,\quad
  \label{eq:min}
\end{eqnarray}
where $n_{\textbf{kq}}$ and $\delta n_{\textbf{kq}}$ are given by the
summands of Eqs.~(\ref{eq:LOFF_neqa}) and (\ref{eq:LOFF_neqb}),
respectively. Furthermore, the FFLO solutions are subject to the
stability condition against phase separation (PS)
\cite{PWY05,Stability,LOFF1},
\begin{equation}
  \frac{\partial^{2}\Omega_{S}}{\partial\Delta^{2}}
  \frac{\partial^{2}\Omega_{S}}{\partial\textbf{q}^{2}}-
  \Big(\frac{\partial^{2}\Omega_{S}}
  {\partial\Delta\partial\textbf{q}}\Big)^{2}>0\,.
\label{eq:sta}
\end{equation}
This condition is equivalent to the positive definiteness of the
particle number susceptibility matrix $\{\partial
n_\sigma/\partial\mu_{\sigma'}\}$ 
% with respect to variations of chemical potentials
\cite{Stability,LOFF1}.  For the Sarma phase (where $\textbf{q}=0$),
Eq.~(\ref{eq:sta}) is reduced to
${\partial^{2}\Omega_{S}}/{\partial\Delta^{2}}>0$ \cite{Stability}.

Equations (\ref{eq:gap}), (\ref{eq:LOFF_neqa}), (\ref{eq:LOFF_neqb})
and (\ref{eq:min}) form a closed set of self-consistent equations,
which can be used to solve for ($\Delta$, $\mu_\uparrow$,
$\mu_\downarrow$, $\textbf{q}$) with various parameters $1/k_{F}a$,
$p$, and $T$, as well as the mass ratio $m_\uparrow/m_\downarrow$, and
obtain the FFLO regions in phase diagrams. Since phase separation
provides an alternative way to accommodate the excessive majority
fermions, some of the mean-field solutions of the FFLO states are
unstable against phase separation. Here we use the stability condition
Eq.~(\ref{eq:sta}) to locate the phase boundary separating stable FFLO
(or Sarma superfluid) phases and the PS phases. As a convention, we
take the heavy (light) species to be spin up (down), and define Fermi
momentum $k_{F}=(3\pi^{2}n)^{1/3}$.  To avoid an artificial jump
across population imbalance $p\equiv \delta n/n=0$ in the phase
diagrams, we take $m=(m_{\uparrow}+m_{\downarrow})/2$ and define the
Fermi temperature as $T_{F}=k_{F}^{2}/2m$ as our energy unit.

Note that for the ($\mathbf{q}=0$) Sarma phases, we will use the pairing
fluctuation theory  described in Ref.~\cite{Guo2009PRA} to determine
the superfluid and pseudogap regions.

\begin{figure}
%  \centerline{\includegraphics[clip,width=3.4in] {FF-Unitary_new.eps}}
  \centerline{\includegraphics[clip,width=3.4in] {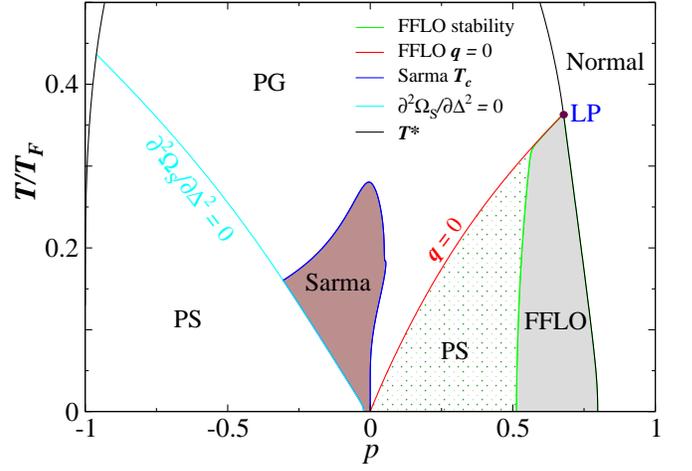}}
  \caption{(Color online) $T$--$p$ phase diagram of a homogeneous
    $^{6}$Li--$^{40}$K mixture at unitarity. Here ``PG" and ``PS"
    indicate pseudogapped normal state and phase separation,
    respectively, and LP labels a Lifshitz point. An FFLO superfluid
    (grey shaded) phase exists in the high $p$ regime when $^{40}$K
    dominates, while it becomes unstable in the dotted region. A
    Sarma superfluid lives in the intermediate $T$ and low $p$ regime
    (brown shaded region). }
\label{fig:Unitary}
\end{figure}

Figure \ref{fig:Unitary} shows the calculated $T$--$p$ phase diagram
for a homogeneous $^{6}$Li--$^{40}$K mixture at unitarity.  Pairing
takes place below the pairing temperature $T^*$ (black solid curve). A
mean-field FFLO solution exists to the lower right of the (red)
$\mathbf{q}=0$ line.  However, stable FFLO states exist only when
$^{40}$K is the majority at relatively high $p$ (in the grey shaded
area). For lower $p$, FFLO states become unstable and phase separation
take places at low $T$ (dotted region), whereas Sarma superfluid
(brown area) and pseudogap states exist at intermediate $T$. The
(green) line that separates the PS and the FFLO phases is given by the
stability condition Eq.~(\ref{eq:sta}).  When $^{6}$Li is the
majority, i.e., $p<0$, phase separation dominates the low $T$
region. Note here that, as we focus on the FFLO phases, we do not
distinguish superfluid and pseudogap states in the PS regions.
Furthor details regarding non-FFLO related phases can be found in
Ref.~\cite{Guo2009PRA}.

\begin{figure}
\centerline{\includegraphics[clip,width=3.4in] {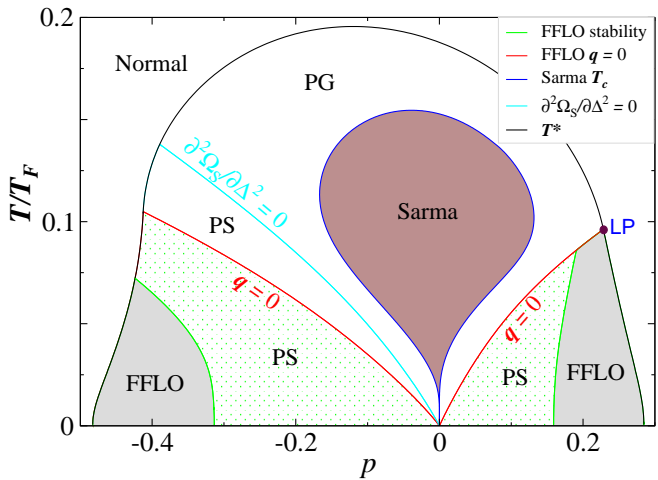}}
\caption{(Color online) $T$--$p$ phase diagram of a homogeneous
  $^{6}$Li--$^{40}$K mixture at $1/k_{F}a=-1$, similar to
  Fig.~\ref{fig:Unitary}.  Here the FFLO phase (gray shaded regions)
  exists for both $p>0$ and $p<0$.}
\label{fig:BCS}
\end{figure}

Shown in Fig. \ref{fig:BCS} is the (near-)BCS counterpart of
Fig.~\ref{fig:Unitary} at $1/k_{F}a=-1$, with much weaker pseudogap
effects.  Here we find stable FFLO phases for $p<0$ as well, when
$^{6}$Li is the majority. This is different from
Refs.~\cite{Stoof09prl,Stoof10pra,Stoof12pra}, which found no LO or
supersolid states in the BCS regime for $p<0$. The $p>0$ part is
rather similar to the unitary case, except that everything moves to
lower $p$ and lower $T$ due to weaker pairing strength. For $p<0$, the
$\mathbf{q}=0$ line splits the PS phase into two regions, representing
unstable Sarma (upper) and FFLO (lower part) phases, respectively.

In both Figs.~\ref{fig:Unitary} and \ref{fig:BCS}, we have found a
Lifshitz point (as labeled ``LP'') within the mean-field treatment,
below which FFLO states emerge.

As the pairing strength grows, the Sarma phase becomes stabilized in a
much larger region, especially for $p<0$ (not shown). However, the
stable FFLO states are squeezed towards very low $T$ and very high $p
\lesssim 1$, and eventually disappear on the BEC side of the Feshbach
resonance. Our result suggests that it is more promising to find FFLO
phases in the unitary regime.

\begin{figure}
%\centerline{\includegraphics[clip,width=3.4in] {diff-mass-FF-Tc-sta_PR_5a.eps}}
\centerline{\includegraphics[clip,width=3.4in] {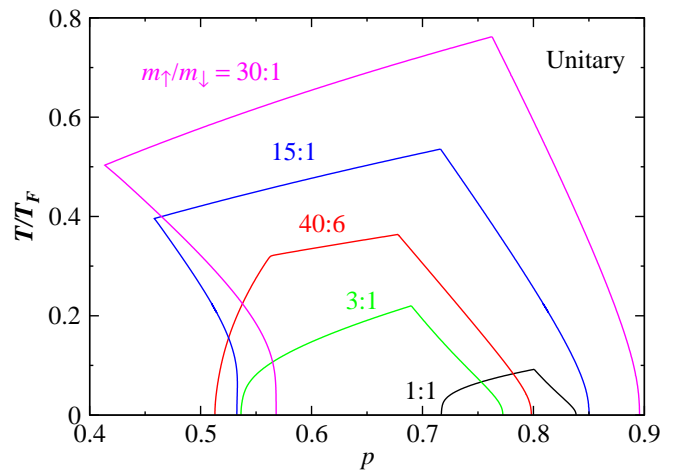}}
\caption{(Color online) $T$--$p$ phase diagram of stable FFLO
  superfluid in Fermi-Fermi mixtures with different mass ratios (as
  labeled) at unitarity.  Large mass ratio enhances FFLO type of
  pairing.}
\label{fig:diff-mass}
\end{figure}

To ascertain the effect of a varying mass ratio
$m_\uparrow/m_\downarrow$, we now focus on the stable FFLO superfluid
phase for $p>0$ at unitarity and compute the phase diagram for a
series of different mass ratios, as shown in
Fig.~\ref{fig:diff-mass}. (The stable FFLO phase for $p<0$ quickly
disappears when $m_\uparrow/m_\downarrow \gtrsim 1.9$).
Now that the mass ratio is changing, it is important to pick the right
energy unit for meaningful comparison. In addition to the $T_F$ used
here, one may alternatively consider using $m_r$ in the definition of
$T_F$. The plot is shown in the supplemental Fig.~S2. However, since
$m_r$ is an average based on the inverse mass, it puts more weight on
the light species, which is more appropriate for the $p<0$ case. For
the large $p>0$ case, where the heavy species dominates, we conclude
that the present definition of $T_F$ is more appropriate.

Figure \ref{fig:diff-mass} suggests that the FFLO $T_c$ increases as
the mass imbalance grows.  At the same time, the phase space of stable
FFLO superfluid also grows much larger as the mass ratio increases. In
comparison with the mass balanced case, i.e.,
$m_\uparrow/m_\downarrow=1$, for $m_\uparrow/m_\downarrow=40:6$, the
enhancement of the FFLO $T_c $ is about 3 times.  For the
$^6$Li--$^{173}$Yb mixture \cite{Fukuhara,Gupta2013PRA} which has a
mass ratio near 30, the enhancement is about 7 times. Such great
enhancement of $T_c$ and enlarged phase space suggest that it is much
easier to find experimentally the exotic FFLO superfluid with a large
mass ratio.
Note that one may also consider measuring $T_c$ in units of the actual
Fermi temperature of the heavy majority atoms, $T_{F,\uparrow} =
k_{F,\uparrow}^2/2m_\uparrow^{}$, which seems to be a natural choice
for Ref.~\protect\cite{Shin08}. In this case, the enhancement of $T_c$
by mass imbalance would be even more dramatic, being 7 and 16 times,
respectively, as shown in the Supplemental Fig.~S1.

\begin{figure}
\centerline{\includegraphics[clip,width=3.3in] {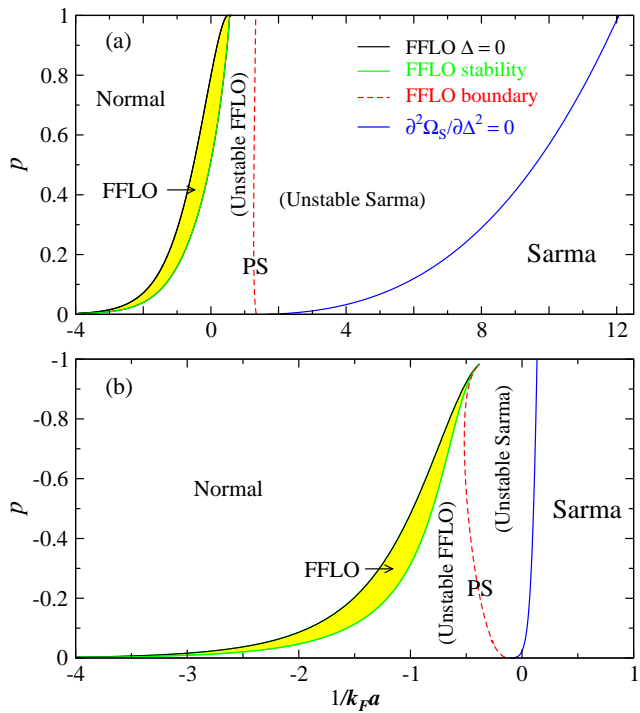}}
\caption{(Color online) Phase diagram of $^{6}$Li--$^{40}$K in the $p$
  -- $1/k_{F}a$ plane at $T=0$ for (a) $p>0$ and (b) $p<0$.  Stable
  FFLO phase lives in the narrow (yellow) shaded regions. Here ``PS''
  labels phase separation (against unstable FFLO and Sarma
  superfluids), divided by the (red) dashed $\mathbf{q}=0$
  line. \vspace*{-2ex} }
\label{fig:ip}
\end{figure}

Shown in Figs.~\ref{fig:ip} are the calculated $p$--$1/k_{F}a$ phase
diagrams of a $^{6}$Li--$^{40}$K mixture at $T=0$ for (a) $p>0$ and
(b) $p<0$, respectively. When $^{40}$K is the majority,
Fig.~\ref{fig:ip}(a) shows that a narrow (yellow shaded) region of
stable FFLO superfluids persists from the BCS through the near-BEC
regime, up to $1/k_Fa \approx 0.55$, as $p$ varies from 0 to
1. Apparently, in the near-BEC regime, the stable FFLO phase exists
only at large $p$.  On the other hand, when $^6$Li is the majority,
the stable FFLO phase moves left completely to the BCS side, as shown
in Fig.~\ref{fig:ip}(b), in agreement with Figs.~\ref{fig:Unitary} and
\ref{fig:BCS}.  In comparison with the equal-mass case \cite{LOFF1},
here the stable FFLO region for $p>0$ is slightly larger, while it
becomes smaller for $p<0$. Here ``PS'' in both figures labels the
regions of FFLO and Sarma superfluids that are unstable against phase
separations.  In both cases, the FFLO $q$ vector increases from 0 in
magnitude as $|p|$ increases along the boundaries of the stable FFLO
phase. The red dashed line separates from unstable FFLO and unstable
Sarma regions.

More details about the behavior of $q$ as a function of interaction
strength and population imbalance $p$ are given in the Supplemental
Information.

It is interesting to note that for $p<0$, due to the left shift of the
PS phase, the stable zero $T$ Sarma superfluid phase has extended into
the unitary regime ($1/k_Fa \gtrsim -0.1$) for small $|p|$, as can
also be seen in Fig.~\ref{fig:Unitary}, where the Sarma phase extends
all the way down to $T=0$ at $p\lesssim 0$. This should be contrasted
with the $p>0$ case and the equal mass case, where zero $T$ Sarma
superfluid can be found only when $1/k_Fa \gtrsim 1.5$ and $1/k_Fa
\gtrsim 0.6$ \cite{Chien06}, respectively.

Since a flat bottom or quasi-uniform trap has been realized
experimentally \cite{GauntPRL2013}, our homogeneous result may be
directly applicable when such a trap is used. 
For a harmonic confining trap which causes inhomogeneity in terms of
population imbalances \cite{ChienPRL,Yip2007PRA76,wang13pra}, we take
the study of the homogeneous case as a necessary first step. In
addition, one may obtain experimentally homogeneous result using a
tomography technique \cite{Shin08}).
In a trap, sandwich-like shell structures will emerge when $p>0$, with
superfluid or pseudogapped normal state in the middle shell
\cite{wang13pra}.  Figure \ref{fig:Unitary} suggests that the FFLO
states may be found locally at low $T$ near the shell interfaces
where one may find suitable population imbalances.

Finally, we note that while more complex crystalline types of FFLO
states are expected to have a lower energy and thus may be found
within the PS phases in our phase diagrams. While one pair of
$\pm\mathbf{q}$ (i.e., the LO state) may lead to a lower energy, two
pairs may further decrease the energy, and so on. It is not obvious
where to stop the sequence. However, we expect this to cause only
minor modifications in our results. Beyond-mean-field treatment may be
more important than higher order crystalline types of pairing. For
example, inclusion of particle-hole fluctuations may lead to a shift
in the location of unitarity \cite{ParticleHoleChannel}.

In summary, our results show that, in order to find the exotic FFLO
states in a 3D Fermi gas, it is most promising to explore Fermi-Fermi
mixtures with a large mass ratio in the unitary regime, where one
expects to see a relatively large phase space volume and a greatly
enhanced superfluid transition temperature when the heavy species is
the majority. While we have focused on the one-plane-wave FFLO, i.e.,
the FF case, such enhancement is present for the LO phase as well,
which has a comparable or lower energy.
These FFLO states may be detected via collective modes
\cite{CooperPRL103}, vortices \cite{AgterbergPRL100}, direct imaging
\cite{Mizushima}, rf spectroscopy \cite{TormaPRL101}, triplet pair
correlations \cite{Demler2012}, and, most directly, by measuring the
pair momentum distribution which should exhibit a peak at a finite
$\mathbf{q}$.
Experimentally, the $T\ll T_{F,\uparrow}$ regime has now been
accessible for $^{6}$Li--$^{40}$K \cite{Shin08}.  With the recent
report of $T\simeq 0.3T_{F,\uparrow}$ \cite{Gupta2013PRA}, it is
hopeful that lower $T$ regime can be accessed for $^{6}$Li--$^{173}$Yb
as well in the near future. These experimental progress makes it
promising to observe the exotic FFLO states if they do exist.

This work is supported by NSF of China (Grant No. 11274267), the
National Basic Research Program of China (Grants No. 2011CB921303 and
No. 2012CB927404), NSF of Zhejiang Province of China (Grant No.
LZ13A040001).

%\vspace*{-2ex}

\bibliographystyle{apsrev} 
\bibliography{Review3}

\begin{thebibliography}{63}
\bibitem[{foo()}]{footnote1}
\bibinfo{note}{Here the supersolid state is just the LO state.}
\end{thebibliography}

\end{document}